\documentclass[12pt]{iopart}
\usepackage{graphicx}
\begin{document}

\title[Scattering of cylindrical cloak]{The scattering of a cylindrical invisibility cloak: reduced parameters and optimization}

\author{L Peng$^{1,2}$, L Ran$^{2}$ and N A Mortensen$^{1}$}

\address{$^1$Department of Photonics Engineering, Technical University of Denmark, DTU-building 345 west, DK-2800 Kongens Lyngby, Denmark\\
$^2$Department of Information and Electronic Engineering, Zhejiang University, Hangzhou 310027, P.R. China}
\ead{plia@fotonik.dtu.dk}
\begin{abstract}
We investigate the scattering of 2D cylindrical invisibility cloaks with simplified constitutive parameters with the assistance of scattering coefficients. We show that the scattering of the cloaks originates not only from the boundary conditions but also from the spatial variation of the component of permittivity/permeability. According to our formulation, we propose some restrictions to the invisibility cloak in order to minimize its scattering after the simplification has taken place. With our theoretical analysis, it is possible to design a simplified cloak by using some peculiar composites like photonic crystals (PCs) which mimic an effective refractive index landscape rather than offering effective constitutives, meanwhile canceling the scattering from the inner and outer boundaries.
\end{abstract}

\maketitle

\section{Introduction}
\label{sec:intro}
Keeping the form of Maxwell equations invariant, coordinate transformation (CT) provides us the possibility to control the electromagnetic field, by changing the distribution of permittivity and permeability, as well as to create an isolated space to hide any objects, i.e. invisibility cloaking~\cite{Pendry:2006,Luo:2008}. In recent years, invisibility cloaking has been attracting significant attention~\cite{Leonhardt:2006,Schurig:2006,Chen:2007,Li:2008,Greenleaf:2008,Alu:2008,Lai:2009,Tretyakov:2009,Liu:2009,Ergin:2010}, due to its novel electromagnetic (EM) characteristics and potential applications.

In general, the invisibility cloaks, realized through a CT approach, possess constitutive parameters being anisotropic and spatially varying. In particular, a 3D spherical invisibility cloak calls for constitutive parameters which are anisotropic and inhomogeneous, but with finite values, which seems realizable in practice~\cite{Pendry:2006}. However, in the 2D case, cylindrical invisibility cloaks commonly possess constitutives not only being anisotropic and spatially dispersive, but also requesting infinite components~\cite{Schurig:2006}, which in practice leaves the realization of any ideal cloak impossible. To overcome this drawback, Schurig {\emph{et al.}} suggested the simplification of the constitutive parameters, i.e. flexibly choosing the constitutives while keeping the spatial distribution of dispersion relations unchanged~\cite{Schurig:2006}, which is based on the ray tracing concept. With such an approach, the first cylindrical cloak shell was successfully constructed.

Since we have to simplify the constitutives of a 2D cylindrical cloak in the practical design and fabrication, the scattering of the cloak itself is unavoidable, i.e. an invisibility cloak with reduced parameters is inherently visible~\cite{YanM:2007}. In order to reduce the scattering of a cloak with reduced parameters, a variety of research work has been done. For instance, a second-order polynomial function was used in the transformation in order to avoid the reflection at the outer boundary~\cite{Cai:2007}, or to minimize the scattering of cloaks with reduced parameters by the optimized simplification~\cite{YanMOE:2007,YanW:2008,Zhang:2010}. Though people have achieved success in the invisibility cloak theory and experiments, a general description of the scattering by a cloak with reduced parameters is unrevealed.

In this paper, we analytically solve the scattering problem of a cylindrical invisibility cloak by applying the classical scattering theorem. With the aid of scattering coefficients, we show the full-wave solution of the scattering by cloaks with ideal as well as reduced parameters. According to our theoretical analysis, the scattering of a cloak with reduced parameters is contributed by both the inner and outer boundary, as well as the spatially varying internal impedance. We point out the way to minimize the scattering of a cloak in practice, in which case its scattering is dominated by the spatially distributed permittivity/permeability rather than the boundary conditions. In addition, the way to realize nearly zero scattering by an approximate cloaking shell is also proposed.

\section{Scattering of cylindrical invisibility cloak}
\label{sec:theory}

Consider the anisotropic and inhomogeneous invisibility cloaks obtained from a CT approach. If the mapping employed in the transformation is simple, i.e. only one coordinate variable is involved in the transformation like the ones proposed in Ref.~\cite{Pendry:2006} or Ref.~\cite{Schurig:2006}, the cloak's $\bar{\bar{\epsilon}}$  and $\bar{\bar{\mu}}$ are diagonal and inhomogeneous. Though the analytical solution of the scattering problem of an ideal cloak can be found by applying the Maxwell equations, the scattered fields by those cloaks with reduced parameters have to be determined numerically~\cite{YanM:2007}. However, the scattering by a multi-layered cloak shell can be found by applying the full wave expansion, satisfying the boundary conditions~\cite{Xi:2009}.

\begin{figure}[b]
\centerline{\includegraphics[width=0.8\columnwidth,draft=false]{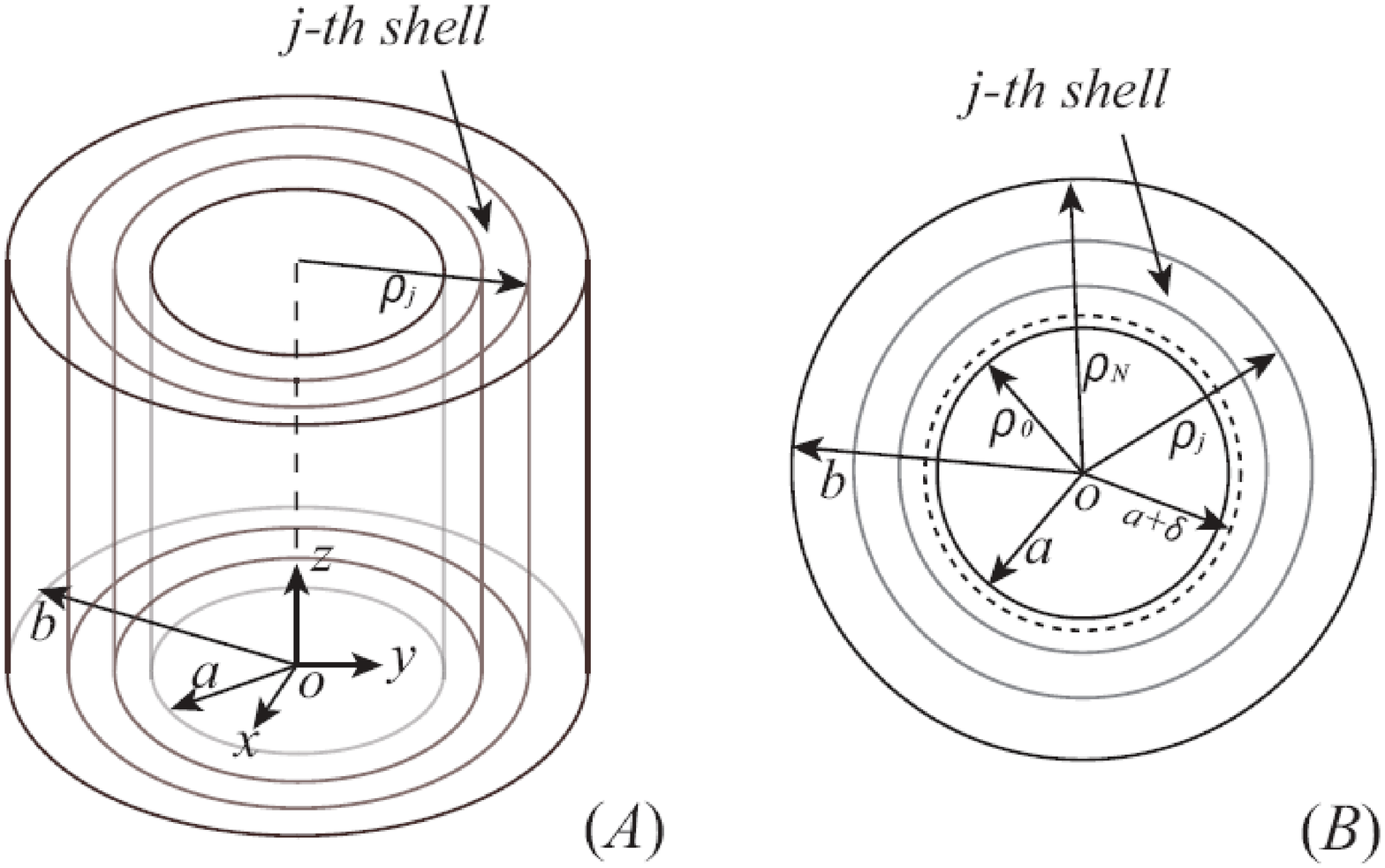}}
\caption{Configuration (A) and Cross section (B) of a multi-layered cylindrical cloak.}
\label{fig1}
\end{figure}

Suppose there is an $N$-layered cloak structure as shown in Fig.~\ref{fig1} (A). The inner and outer radius of this cloak shell are $a$ and $b$, respectively. $\rho_j$ is the outer radius of the  $j$-th shell,  $\rho_0=a$ stands for the radius of the inner region,  $\rho_N=b$ is the outer radius. Inside the  $j$-th layer with thickness  $d_j=\rho_j-\rho_{j-1}$, see Fig.~\ref{fig1}, the permittivity and the permeability are constant and denoted by  $\bar{\bar{\epsilon}}_j$  and $\bar{\bar{\mu}}_j$ . If $\bar{\bar{\epsilon}}_j$ and $\bar{\bar{\mu}}_j$ vary continuously in space, then a continuous cloak shell can be obtained if we let $N\rightarrow\infty$  and  $\max{d_j}\rightarrow 0$ ($j=1,2,3...N$ ). For simplicity, the permittivity and permeability in each layer of the multi-layered cloak are assumed to be anisotropic but diagonal, e.g. $\bar{\bar{\epsilon}}_j/\epsilon_0={\rm diag}\left[\epsilon^r_{\rho,j},\epsilon^r_{\phi,j},\epsilon^r_{z,j}\right]$. Here, we consider the TM polarization case, i.e. the electric field is polarized along $z$ direction, while the same procedure can be applied to a TE polarization case.

Basically, the field inside each layer can be easily expressed by a superposition of vector cylindrical waves~\cite{WChew:1995,Xi:2009}. We rewrite the field inside the $j$-th layer as
\begin{equation}
\label{eq:efield}
E_{z,j}=\sum_{m=-\infty}^{\infty}[C_{1,m}^jJ_{\nu_j}(k_{\rho,j}\rho)+C_{2,m}^jH_{\nu_j}^{(1)}(k_{\rho,j}\rho)]e^{im\phi},
\end{equation}
with $\nu_j=m(k_{\rho,j}/k_{\phi,j})$, $k_{\rho,j}=k_0\sqrt{\epsilon^r_{z,j}\mu^r_{\phi,j}}$, $k_{\phi,j}=k_0\sqrt{\epsilon^r_{z,j}\mu^r_{\rho,j}}$, and $k_0=\omega\sqrt{\epsilon_0\mu_0}$. Here, $C_{1,m}^j$ and $C_{2,m}^j$ are the unknown expansion coefficients inside the $j$-th layer which need to be determined by matching the boundary conditions. In the outer space, $C_{1,m}^{N+1}$ and $C_{2,m}^{N+1}$ are used to represent the total field in the outer free space, composed of the incident and the scattered fields. In Eq.~(\ref{eq:efield}), the first term represents the standing wave, while the second one is the scattered wave in each layer~\cite{WChew:1995}. Boundary conditions are applied to determine the unknown coefficients. At the interface between the $j$-th and the $j+1$-th layers, tangential electric and magnetic fields should be continuous, hence we find
\begin{equation}
\label{eq:matrix}
\left[\begin{array}{c}
C_{1,m}^{j+1}\\C_{2,m}^{j+1}\end{array}\right]
={\bar{\bar{P}}_{m,j+1}}^{-1}\bar{\bar{Q}}_{m,j}\left[\begin{array}{c}C_{1,m}^j\\C_{2,m}^j\end{array}\right],
\end{equation}
with
\begin{equation}
\label{eq:t1}
\bar{\bar{P}}_{m,j+1}=\left[\begin{array}{cc}J_{\nu_{j+1}}(k_{\rho,{j+1}}\rho_j)&H_{\nu_{j+1}}^{(1)}(k_{\rho,{j+1}}\rho_j)\\\frac{k_{\rho,{j+1}}}{\mu^r_{\phi,{j+1}}}J_{\nu_{j+1}}'(k_{\rho,{j+1}}\rho_j)&\frac{k_{\rho,{j+1}}}{\mu^r_{\phi,{j+1}}}{H_{\nu_{j+1}}^{(1)}}'(k_{\rho,{j+1}}\rho_j)\end{array}\right],
\end{equation}
\begin{equation}
\label{eq:t2}
\bar{\bar{Q}}_{m,j}=\left[\begin{array}{cc}J_{\nu_{j}}(k_{\rho,{j}}\rho_j)&H_{\nu_{j}}^{(1)}(k_{\rho,{j}}\rho_j)\\\frac{k_{\rho,{j}}}{\mu^r_{\phi,{j}}}J_{\nu_{j}}'(k_{\rho,{j}}\rho_j)&\frac{k_{\rho,{j}}}{\mu^r_{\phi,{j}}}{H_{\nu_{j}}^{(1)}}'(k_{\rho,{j}}\rho_j)\end{array}\right].
\end{equation}

Similarly, we consider the boundary conditions at any other interface, and the two unknown coefficients in the free space can be evaluated by
\begin{equation}
\label{eq:matrixfull}
\left[\begin{array}{c}
P_{m,N+1}\\Q_{m,N+1}\end{array}\right]
=\prod_{j=1}^{N}{\bar{\bar{P}}_{m,j+1}}^{-1}\bar{\bar{Q}}_{m,j}\left[\begin{array}{c}C_{1,m}^1\\C_{2,m}^1\end{array}\right].
\end{equation}

Equation~(\ref{eq:matrixfull}) is the common relationship that a multi-layered cloak should satisfy, with which the scattered field of a cloak consisting of several layers can be conveniently worked out~\cite{Xi:2009}. However, since the 2D invisibility cloak requires that $k_{\phi,j}$ tends to zero near the inner interface, then $\nu_j$ is singular which makes both $\bar{\bar{T}}_{1,m}$ and $\bar{\bar{T}}_{2,m}$ being ill-conditioned and the numerical evaluation of Eq.~(\ref{eq:matrixfull}) fails.

However, since our purpose is to find out the scattered field, the scattering coefficients are of course very helpful and simplifying the formulation may also overcome the drawback of Eq.~(\ref{eq:matrixfull}). The scattering coefficient for the $m$-th cylindrical wave in the $j$-th layer is defined as $R_{m}^j=C_{2,m}^j/C_{1,m}^j$. From Eq.~(\ref{eq:matrix}), we easily arrive at
\begin{equation}
\label{eq:scattceff}
R_m^{j+1}=\frac{A_m^j+B_m^jR_m^j}{C_m^j+D_m^jR_m^j},
\end{equation}
where $A_m^j$, $B_m^j$, $C_m^j$, and $D_m^j$ are functions of $\nu$, $k_\rho$, $k_\phi$ in both the $j$-th and $j+1$-th layers. If we further assume that all the layers are very thin, and let $N\rightarrow\infty$ and $\max{d_j}\rightarrow 0$, i. e. the cloak structure gradually tends to be continuous, we can rewrite $R_m^{j+1}$, $R_m^{j}$, $A_m^j$, $B_m^j$, $C_m^j$ and $D_m^j$ as $R_m+\Delta R_m$, $R_m$, $A_m+\Delta A_m$, $B_m+\Delta B_m$, $C_m+\Delta C_m$ and $D_m+\Delta D_m$, respectively. Hence we find
\begin{equation}
\label{eq:sceffdif}
\fl \Delta R_m=\frac{A_m+\Delta A_m+(B_m-C_m+\Delta B_m-\Delta C_m)R_m-(D_m+\Delta D_m){R_m}^2}{C_m+\Delta C_m+(D_m+\Delta D_m)R_m}.
\end{equation}
Balancing the first order differentials in Eq.~(\ref{eq:sceffdif}), we get
\begin{equation}
\label{eq:riccati}
R'_m(\rho)=\frac{\pi \rho}{2i}\left[A'_m+(B'_m-C'_m)R_m(\rho)-D'_mR^2_m(\rho)\right],
\end{equation}
with
\begin{eqnarray}
\fl A'_m=-J_{\nu}(k_{\rho}\rho)\left[k_{\rho} k'_{\rho}\rho J''_{\nu}(k_{\rho}\rho)+k'_{\rho}J'_{\nu}(k_{\rho}\rho)+k_\rho \nu' \frac{\partial}{\partial\nu}J'_{\nu}(k_{\rho}\rho)\right]\nonumber\\
+J'_{\nu}(k_{\rho}\rho)\left[k_{\rho} k'_{\rho}\rho J'_{\nu}(k_{\rho}\rho)+k_{\rho}\frac{\mu'_\phi}{\mu_\phi}J_{\nu}(k_{\rho}\rho)+k_\rho \nu' \frac{\partial}{\partial\nu}J_{\nu}(k_{\rho}\rho)\right],\nonumber
\end{eqnarray}
\begin{eqnarray}
\fl B'_m=-H^{(1)}_{\nu}(k_{\rho}\rho)\left[k_{\rho} k'_{\rho}\rho J''_{\nu}(k_{\rho}\rho)+k'_{\rho}J'_{\nu}(k_{\rho}\rho)+k_\rho \nu' \frac{\partial}{\partial\nu}J'_{\nu}(k_{\rho}\rho)\right]\nonumber\\
+{H^{(1)}_{\nu}}'(k_{\rho}\rho)\left[k_{\rho} k'_{\rho}\rho J'_{\nu}(k_{\rho}\rho)+k_{\rho}\frac{\mu'_\phi}{\mu_\phi}J_{\nu}(k_{\rho}\rho)+k_\rho \nu' \frac{\partial}{\partial\nu}J_{\nu}(k_{\rho}\rho)\right],\nonumber
\end{eqnarray}
\begin{eqnarray}
\fl C'_m=J_{\nu}(k_{\rho}\rho)\left[k_{\rho} k'_{\rho}\rho {H^{(1)}_{\nu}}''(k_{\rho}\rho)+k'_{\rho}{H^{(1)}_{\nu}}'(k_{\rho}\rho)+k_\rho \nu' \frac{\partial}{\partial\nu}{H^{(1)}_{\nu}}'(k_{\rho}\rho)\right]\nonumber\\
-J'_{\nu}(k_{\rho}\rho)\left[k_{\rho} k'_{\rho}\rho {H^{(1)}_{\nu}}'(k_{\rho}\rho)+k_{\rho}\frac{\mu'_\phi}{\mu_\phi}{H^{(1)}_{\nu}}(k_{\rho}\rho)+k_\rho \nu' \frac{\partial}{\partial\nu}{H^{(1)}_{\nu}}(k_{\rho}\rho)\right],\nonumber
\end{eqnarray}
\begin{eqnarray}
\fl D'_m={H^{(1)}_{\nu}}(k_{\rho}\rho)\left[k_{\rho} k'_{\rho}\rho {H^{(1)}_{\nu}}''(k_{\rho}\rho)+k'_{\rho}{H^{(1)}_{\nu}}'(k_{\rho}\rho)+k_\rho \nu' \frac{\partial}{\partial\nu}{H^{(1)}_{\nu}}'(k_{\rho}\rho)\right]\nonumber\\
-{H^{(1)}_{\nu}}'(k_{\rho}\rho)\left[k_{\rho} k'_{\rho}\rho {H^{(1)}_{\nu}}'(k_{\rho}\rho)+k_{\rho}\frac{\mu'_\phi}{\mu_\phi}{H^{(1)}_{\nu}}(k_{\rho}\rho)+k_\rho \nu' \frac{\partial}{\partial\nu}{H^{(1)}_{\nu}}(k_{\rho}\rho)\right],\nonumber
\end{eqnarray}
For more details of the technical work, please refer to the appendix. Containing the term involving $R^2_m(\rho)$, Eq.~(\ref{eq:riccati}) becomes the non-linear Riccati equation for continuous invisibility cloaks. It could be solved numerically with initial value at the boundaries.

To find out the scattering coefficients in the outer free space, we should find out the scattering coefficients on the inner interface first. For the ideal cloaks, the permittivity/permeability on the inner boundary is infinite. This prevents us from the evaluation of the scattering coefficients, hence we have to assume that the cloak's inner radius has a small perturbation $\delta$, i. e. the inner radius is now $a+\delta$ but the rest of the parameters are kept unchanged. Notice that there is no scattered field inside the inner region ($\rho<a+\delta$), then the scattering coefficients on the new inner boundary can be evaluated by applying Eq.~(\ref{eq:matrix})
\begin{equation}
\label{eq:initial}
\fl R_m(a+\delta)=\frac{\eta_{a+\delta}J_m[k_i(a+\delta)]J'_{\nu_{a+\delta}}[k_{a+\delta,\rho}(a+\delta)]-J'_m[k_i(a+\delta)]J_{\nu_{a+\delta}}[k_{a+\delta,\rho}(a+\delta)]}{\eta_{a+\delta}J_m[k_i(a+\delta)]{H^{(1)}_{\nu_{a+\delta}}}'[k_{a+\delta,\rho}(a+\delta)]-J'_m[k_i(a+\delta)]{H^{(1)}_{\nu_{a+\delta}}}[k_{a+\delta,\rho}(a+\delta)]},
\end{equation}
with $\eta_{a+\delta}=\sqrt{{\epsilon_{a+\delta,z}}/{\mu_{a+\delta,\phi}}}$ and $k_i$ being the wave number in the region $\rho<a+\delta$. The initial scattering coefficients of any cloak structure can be obtained by letting $\delta\rightarrow 0$ in Eq.~(\ref{eq:initial}). For example, for ideal cloaks, ${\frac{k_{a+\delta,\rho}}{\mu_{a+\delta,\phi}}}\rightarrow 0$ and ${\nu_{a+\delta}}\rightarrow 0$ lead to $R_0(a+\delta)\rightarrow -\frac{J_0(k_{a,\rho}a)}{H_0^{(1)}(k_{a,\rho}a)}$ and $R_m(a+\delta)\rightarrow 0 (m\neq 0)$.

\begin{figure}[b]
\centerline{\includegraphics[width=0.8\columnwidth,draft=false]{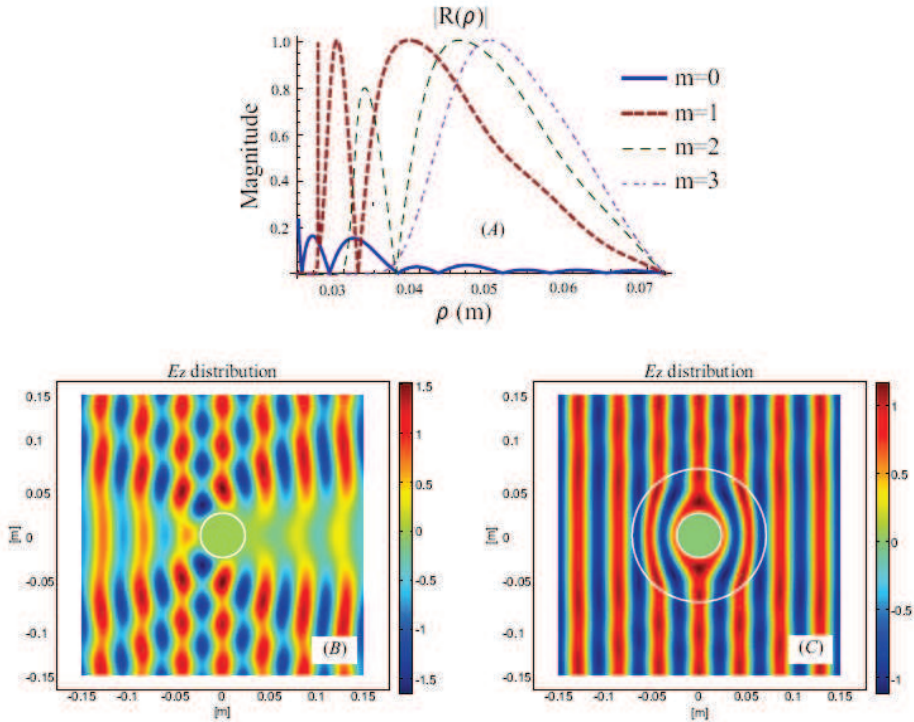}}
\caption{(Color online) (A) The absolute scattering coefficients for an ideal cloak with linear transformation function. (B) The total near field distribution with a bare perfect electric conducting (PEC) cylinder scattering. (C) The near field distribution with a PEC cylinder covered by an ideal cloak. In the calculation, operation frequency is 7GHz.}
\label{fig2}
\end{figure}

To verify our theoretical formulation, here we assume there is a cylindrical invisibility cloak with $a$=0.024m, $b$=0.072m and its constitutive parameters $\bar{\bar{\epsilon}}^r=\bar{\bar{\mu}}^r={\rm diag}\left[\frac{\rho-a}{\rho},\frac{\rho}{\rho-a},(\frac{b}{b-a})^2(\frac{\rho-a}{\rho})\right] $. By using Eqs.~(\ref{eq:riccati}) and~(\ref{eq:initial}), the scattering coefficients can be conveniently evaluated. Fig.~\ref{fig2} (A) shows the scattering coefficients for the lowest four cylindrical waves ($m$=0,1,2,3).

From Fig.~\ref{fig2}, we see that all the scattering coefficients ($|R_m|$) are zero on the outer boundary. Since an ideal cloak perfectly matches the outer space, no scattered wave will be excited. Likewise, the inner boundary, $|R_m|$ is also zero, except the 0-th order case. Though non-zero $R$, no electric field can be excited inside the inner region~\cite{Pendry:2006}, which can also be proved by applying Eq.~(\ref{eq:matrix}). In Fig.~\ref{fig2} (A), $R_0$ varies fast near the inner boundary, hence the total scattered field is sensitive to any perturbation of the inner boundary, as reported in Ref.~\cite{Ruan:2007}.

\section{Cloaks with simplified parameters}
\label{sec:scsimplified}
In the previous section, we have obtained the scattering coefficients for an arbitrary invisibility cloak structure with cylindrical symmetry. However, the permittivity and permeability of ideal cloaks are not only anisotropic and inhomogeneous, but also have infinite components ($\epsilon_\phi$ and $\mu_\phi$) on the inner boundary~\cite{Schurig:2006}. Hence, in order to realize the practical design and fabrication, we have to simplify the constitutive parameters, i. e. maintain the dispersion relation distribution, while releasing our choice of $\epsilon$ and $\mu$ to a more realistic set~\cite{Schurig:2006,CaiNP:2007,Cai:2007,YanMOE:2007}. Unfortunately, this simplification generally affects the scattering coefficients, i.e. the mismatched inner and outer boundaries, as well as the variation of $\mu_\phi$ (for TM case), will cause non-zero scattering coefficients, which we see below in detail.

In Ref.~\cite{Cai:2007}, to minimize the scattered field after the simplification, high order coordinate transformation was applied, for the purpose of impedance matching on the boundary $\rho=b$. While the inner boundary is usually assumed to be perfectly conducting~\cite{Chen:2007,Cai:2007,Xi:2009}, by doing this the inner region is isolated from the outer space. However, the 0-th order cylindrical wave can still see the inner region and detect the perfect conducting wall~\cite{YanM:2007}. Although with the help of some numerical method like genetic algorithm (GA) we can optimize the scattered field by a multi-layered cloak~\cite{Xi:2009}, the physical interpretation of the scattering is missing. Returning to Eq.~(\ref{eq:initial}), the initial value of $R$ is commonly determined by both the cloak and the object placed inside. However, we notice that $R(a)$ can be independent of the permeability and keeps zero if $k_{a,\rho}=0$, which means that the scattering of a cloak will be irrelevant to the inner objects. Again, in Eq.~(\ref{eq:riccati}), the alternative $\mu_\phi$ after the simplification may introduce the disturbance of scattering coefficients through the term containing $\mu'_\phi/\mu_\phi$. To minimize the contribution from this term, the other terms should dominate, i.e. $k'_\rho$ can not be zero.

If the coordinate transformation is $\rho'=f(\rho)$, $\phi'=\phi$ and $z'=z$, all the above restrictions can be expressed as $f'(b)=1$, $f'(a)=0$, $f''(\rho)\neq 0$, with natural conditions $f(a)=0$ and $f(b)=b$. Any functions satisfying these restrictions can be applied to the cloak design. Here, a third-order polynomial function is adopted, i.e. $\rho'=A\rho^3+B\rho^2+C\rho+D$ with $A=-\frac{a+b}{(b-a)^3}$, $B=\frac{1}{2(b-a)}-\frac{3(a+b)}{2}A$, $C=1-3Ab^2-2bB$, and $D=-(Aa^3+Ba^2+Ca)$. We should notice that the choice for higher-ordered polynomial function is not unique, which indicates that the optimization might be accomplished by using some numerical method like GA.

Now we address some numerical calculations to verify our theoretical analysis. We suppose an infinitely long cylinder with radius $r$=24mm made of PEC is placed in free space. The scattering coefficients of this PEC cylinder under TM polarized wave illuminating are listed in the second column of Tab.~\ref{table1}, while the near field distribution is shown in Fig.~\ref{fig2} (B).

\begin{figure}[b]
\centerline{\includegraphics[width=0.8\columnwidth,draft=false]{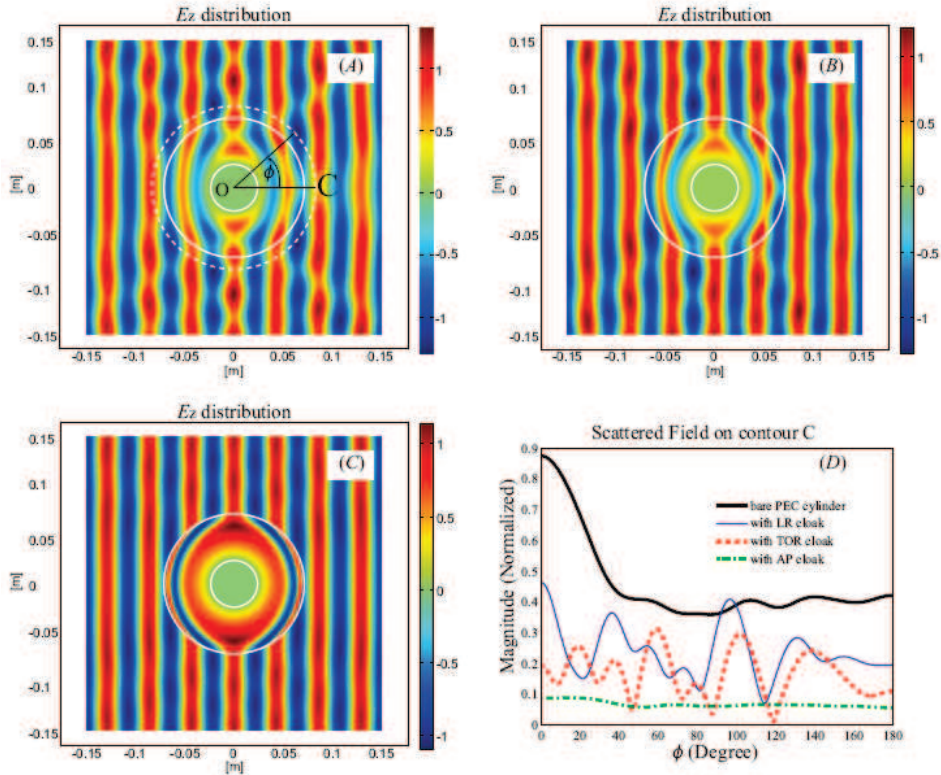}}
\caption{(Color online) Total electric field distribution for a PEC cylinder covered by (A) a simplified cloak coming from a linear CT; (B) a simplified cloak coming from a third-order polynomial CT; (C) an approximate cloak shell. (D) The magnitude of the scattered electric field on the virtue contour C ($\rho$=0.084m). Due to the symmetry, here we only show the distribution in $0<\phi<\pi$. In all of the calculation, $f$=7GHz.}
\label{fig3}
\end{figure}

Next we assume that a cylindrical cloak with $a$=24mm and $b$=72mm is applied to hide the specified PEC cylinder. First, we apply a linear function to complete the CT, as previously stated. In practice, the constitutive parameters have to be simplified. Here we make $\mu_\phi$=1 while we tune the other parameters correspondingly~\cite{Schurig:2006}. The near field distribution of the scattering by this simplified cloak is shown in Fig.~\ref{fig3} (A), while the scattering coefficients for the lowest four order cylindrical waves are listed in the third column in Tab.~\ref{table1}. We clearly find that the non-zero scattering coefficients make the cloak visible~\cite{YanM:2007}. Comparing the second and the third columns in Tab.~\ref{table1}, we see that though with non-zero scattering coefficients, this simplified cloak has a larger size but less scattering than the PEC cylinder.
\begin{table}
\caption{\label{table1}The absolute scattering coefficients for the lowest four cylindrical waves are shown. In the calculation, $f$=7GHz. $|R_c|$: the bare PEC cylinder case; $|R_l|$: PEC cylinder covered by a simplified cloak from linear CT; $|R_{3ord}|$: PEC cylinder covered by a simplified cloak from a third-order polynomial CT; $|R_{app}|$: PEC cylinder covered by an approximate cloak shell.}
\begin{indented}
\item[]\begin{tabular}{@{}lllll}
\br
$m$&$|R_c|$&$|R_l|$&$|R_{3ord}|$&$|R_{app}|$\\
\mr
0&	0.9036&	0.801&	0.354&	0.1187\\
1&	0.3004&	0.197&	0.135&	0.0068\\
2&	0.9934&	0.082&	0.031&	0.0066\\
3&	0.7418&	0.245&	0.060&	0.0120\\
\br
\end{tabular}
\end{indented}
\end{table}

Second, a third-order polynomial function, which can be uniquely determined by the above analysis, is applied in the CT. Though the cloak still needs simplification, its scattering is only caused by the alternated $\mu_\phi$, see Eq.~(\ref{eq:riccati}), hence we would expect the reduction of all the scattering coefficients. We still make $\mu_\phi$=1 after the simplification. The total field distribution is shown in Fig.~\ref{fig3} (B), while the lowest four scattering coefficients are listed in the fourth column in Tab.~\ref{table1}. We see that though the scattering coefficients are non-zero again, this cloak scatters less EM power than the one shown in Fig.~\ref{fig3} (A).

From Fig.~\ref{fig3} and Tab.~\ref{table1}, a simplified cloak structure related to a third-order polynomial CT has less scattering than the one related to a linear CT. However the scattering contributed from the $\mu_\phi$ term in Eq.~(\ref{eq:riccati}) is still unavoidable. In fact, by applying high-order polynomial transformation functions and some numerical optimization, the cloak's parameters can be optimized to minimize the scattering involved by the $\mu_\phi$ term. Here, we point out one possible analytical solution.

We go to Eq.~(\ref{eq:riccati}) and if the term containing $\mu'_\phi/\mu_\phi$ is zero, then the scattering of the cloak is dominated by the spatially dispersive refractive index. For this purpose, we have $\mu'_\phi/\mu_\phi=0$, or $\rho'=b^{1-X}\rho^X$, with $X$ being determined by boundary conditions. Rigorously speaking, functions of this kind can not become the transformation function since $\rho'\neq 0$ if $\rho\neq 0$. However, properly choosing the two unknown coefficients can still support an approximate cloak shell~\cite{Cummer:2009}. An approximate cloak shell coming from this kind of transformation function will possess constitutives as $\bar{\bar{\epsilon}}^r=\bar{\bar{\mu}}^r=diag\left[X^{-1},X,Xb^{2-2X}\rho^{2C-2}\right]$, i.e. no singularity existing in both the permittivity and the permeability. Since this kind of invisibility cloak shell has been reported in Ref.~\cite{Cummer:2009}, we do not place too much discussion here. The scattering coefficients are summarized in the last column in Tab.~\ref{table1} and the electric field distribution is given in Fig.~\ref{fig3} (C). We see that the approximate cloak shell scatters little EM power, which in principle "shrinks" the non-zero scatter~\cite{Cummer:2009}.

\section{Conclusion}
\label{sec:concl}
In conclusion, we have solved the scattering problem of 2D cloak structures with cylindrical symmetry, for both the ideal and simplified cases. We show that the scattering coefficients of a cloak are determined by both the inhomogeneous refractive index and the inhomogeneous $\mu_\phi$($\epsilon_\phi$), as well as the inner and outer boundary conditions. In the practical fabrication, to avoid the infinities of an ideal cloak, the cloak's constitutive parameters have to be simplified, which commonly makes an otherwise invisible cloak visible.

We point out the way to avoid the scattering from both the outer and the inner boundaries, i.e. to have the outer boundary matched to the free space and make the refractive index tend to zero to minimize the scattering from the inner boundary, e.g. high-order polynomial CT can be applied to the cloak design. However, though numerical results show the reduction of scattering, the third-order polynomial based CT can not guarantee the zero scattering due to the changing of $\mu_\phi$ after the simplification. Some numerical method like GA can be adopted to further optimize the scattering of a simplified cloak. Our theoretical work shown in this paper clearly exhibits the role of the inhomogeneous refractive index. With minimizing the scattering involved by the $\mu_\phi$ term and the two boundaries, dielectric metamaterials and photonic crystals, which prefer effective spatial dispersion rather than effective constitutive parameters, could be conveniently applied to construct an invisibility cloak~\cite{Peng:2007,Peng:2010,Yaroslav:2010}, in which case the topology optimization may stimulate the practical design and fabrication~\cite{Ole:2009}. Our theoretical work in this paper would improve the practical fabrication and application of invisibility cloaks in future.

\ack This work was financially supported by the \emph{Danish National Advanced Technology Foundation} (No. 004-2007-1), the \emph{Sino-Danish Scientific and Technological Cooperation, Danish Agency for Science, Technology and Innovation} (No. 09-075623), the \emph{National Science Foundation of China (NSFC)} (No. 61071063), \emph{863 Project} (No. 2009AA01Z227) and NCET-07-0750.

\appendix
\section{The derivation of Eq.~(\ref{eq:riccati})}
We rewrite Eq.~(\ref{eq:matrix}) as
\begin{equation}
\label{eq:a1}
\left[\begin{array}{c}
C_{1,m}^{j+1}\\C_{2,m}^{j+1}\end{array}\right]
=\frac{\pi}{2i}\left[\begin{array}{cc}C_m^j&D_m^j\\A_m^j&B_m^j\end{array}\right]\left[\begin{array}{c}C_{1,m}^j\\C_{2,m}^j\end{array}\right],
\end{equation}
and after some algebra operation we find
\begin{eqnarray}
\fl A_m^j=-k_{\rho,j+1}\rho_jJ'_{\nu_{j+1}}(k_{\rho,j+1}\rho_j)J_{\nu_j}(k_{\rho,j}\rho_j)
+k_{\rho,j}\rho_j\frac{\mu_{\phi,j+1}^r}{\mu_{\phi,j}^r}J_{\nu_{j+1}}(k_{\rho,j+1}\rho_j)J'_{\nu_j}(k_{\rho,j}\rho_j),\nonumber\\
\fl B_m^j=-k_{\rho,j+1}\rho_jJ'_{\nu_{j+1}}(k_{\rho,j+1}\rho_j)H^{(1)}_{\nu_j}(k_{\rho,j}\rho_j)
+k_{\rho,j}\rho_j\frac{\mu_{\phi,j+1}^r}{\mu_{\phi,j}^r}J_{\nu_{j+1}}(k_{\rho,j+1}\rho_j){H^{(1)}_{\nu_j}}'(k_{\rho,j}\rho_j),\nonumber\\
\fl C_m^j=k_{\rho,j+1}\rho_j{H^{(1)}_{\nu_{j+1}}}'(k_{\rho,j+1}\rho_j)J_{\nu_j}(k_{\rho,j}\rho_j)
-k_{\rho,j}\rho_j\frac{\mu_{\phi,j+1}^r}{\mu_{\phi,j}^r}H^{(1)}_{\nu_{j+1}}(k_{\rho,j+1}\rho_j)J'_{\nu_j}(k_{\rho,j}\rho_j),\nonumber\\
\fl D_m^j=k_{\rho,j+1}\rho_j{H^{(1)}_{\nu_{j+1}}}'(k_{\rho,j+1}\rho_j)H^{(1)}_{\nu_j}(k_{\rho,j}\rho_j)
-k_{\rho,j}\rho_j\frac{\mu_{\phi,j+1}^r}{\mu_{\phi,j}^r}H^{(1)}_{\nu_{j+1}}(k_{\rho,j+1}\rho_j){H^{(1)}_{\nu_j}}'(k_{\rho,j}\rho_j).\nonumber
\end{eqnarray}
From Eq.~(\ref{eq:a1}), we can derive Eq.~(\ref{eq:scattceff}). Furthermore, assuming that all the cylindrical shells are thin, i.e. $\max{d_j}\rightarrow 0$, we can rewrite $R_m^{j+1}$, $R_m^{j}$, $A_m^j$, $B_m^j$, $C_m^j$ and $D_m^j$ as $R_m+\Delta R_m$, $R_m$, $A_m+\Delta A_m$, $B_m+\Delta B_m$, $C_m+\Delta C_m$ and $D_m+\Delta D_m$. The $\Delta$ sign appears in those terms involving $\Delta\rho=\rho_{j+1}-\rho_{j}$. Thus Eq.~(\ref{eq:scattceff}) has a new form as Eq.~(\ref{eq:sceffdif}) which is rewritten here
\begin{equation}
\label{eq:a2}
\fl \Delta R_m=\frac{A_m+\Delta A_m+(B_m-C_m+\Delta B_m-\Delta C_m)R_m-(D_m+\Delta D_m){R_m}^2}{C_m+\Delta C_m+(D_m+\Delta D_m)R_m},
\end{equation}
with
\begin{eqnarray}
\label{eq:a3}
A_m&=&0,\nonumber\\
B_m&=&\frac{2i}{\pi},\nonumber\\
C_m&=&\frac{2i}{\pi},\nonumber\\
D_m&=&0.
\end{eqnarray}
In deriving Eq.~(\ref{eq:a3}), we use the Wronskian of Hankel functions~\cite{Abramowitz:1965}
\begin{equation}
\label{eq:wronskian}
H_v^{(1)}(x)J'_v(x)-J_v(x){H_v^{(1)}}'(x)=-\frac{2i}{\pi x}.
\end{equation}
Furthermore, simplifying Eq.~(\ref{eq:a2}), we get
\begin{equation}
\label{eq:a4}
\Delta R_m(C_m+\Delta C_m+\Delta D_mR_m)=\Delta A_m+(\Delta B_m-\Delta C_m)R_m-\Delta D_m{R_m}^2.
\end{equation}
Balancing the first order differentials in Eq.~(\ref{eq:a4}), we can get Eq.~(\ref{eq:riccati}).

\end{document}